Waiting, Banning, and Embracing: An Empirical Analysis of Adapting Policies for Generative AI in Higher Education

**Ping Xiao**
Melbourne Business School
University of Melbourne
p.xiao@mbs.edu

**Yuanyuan Chen**
The Culverhouse College of Business
The University of Alabama
ychen200@cba.ua.edu

**Weining Bao**
School of Business
University of Connecticut
weining.bao@uconn.edu

This Version: May 24th, 2023

Waiting, Banning, and Embracing: An Empirical Analysis of Adapting Policies for Generative AI in Higher Education

**Abstract**

Generative AI tools such as ChatGPT have recently gained significant attention in higher education. This study aims to understand how universities establish policies regarding the use of AI tools and explore the factors that influence their decisions. Our study examines ChatGPT policies implemented at universities around the world, including their existence, content (i.e., whether they embrace or ban ChatGPT and their coverage) and issuance dates. Specifically, we analyzed the top 500 universities according to the 2022 Quacquarelli Symonds (QS) World University Rankings. Our findings indicate that there is significant variation in university ChatGPT policies. Less than one-third of the universities included in the study had implemented ChatGPT policies. Of the universities with ChatGPT policies, approximately 67.4% embraced ChatGPT in teaching and learning, more than twice the number of universities that banned it. The majority of the universities that ban the use of ChatGPT in assessments allow individual instructors to deviate from this restrictive policy. Our empirical analysis identifies several factors that are significantly and positively correlated with a university's likelihood of having a ChatGPT policy, including the university's academic reputation score, being located in an English-speaking country, and the general public and social media's attitudes toward "ChatGPT." In addition, we found that a university's likelihood of having a ban policy is positively associated with faculty-student ratio, citations, and the English-speaking country dummy, while negatively associated with the number of peer universities within the same country that have banned ChatGPT. We discuss the challenges faced by universities based our empirical findings.

1. **Introduction**

Recent years have witnessed remarkable advancements in generative artificial intelligence (AI) owing to the progress in machine learning techniques. These breakthroughs have led to the development of AI-powered technologies that are revolutionizing numerous industries, including education. Among these innovations is ChatGPT, a natural language processing (NLP) chatbot developed by OpenAI. ChatGPT can engage in human-like conversations, answer questions, and generate sophisticated text, code, and creative content that are indistinguishable from that produced by a human (Dwivedi et al., 2023). Throughout history, the education sector has experienced transformative impacts from modern technologies like computers, mobile phones, and the Internet. However, none have had such a profound and rapid impact as ChatGPT. Ever since its release in November 2022, ChatGPT has garnered considerable attention in higher education due to its remarkable productivity and effectiveness, its ability to learn continuously, and its free accessibility.[1]

Educators hold divergent opinions regarding generative AI tools. Some educators are concerned that these tools, capable of quickly generating substantial amounts of readable, well-crafted, and unique text, may foster academic dishonesty and provide unfair assistance in programming and problem-solving endeavours (Sullivan et al. 2023). Multiple reports have demonstrated ChatGPT's capabilities, providing notable examples of its success in passing graduate-level exams in law, business, and even the U.S. Medical Licensing Exam.[2] The higher education sector has long been grappled with the problem of contract cheating, and the advent of generative AI tools exacerbates this problem by significantly reducing the costs associated with cheating. In addition, there are concerns that students will lack creativity and critical thinking skills and will be manipulated by AI models. Conversely, other educators are less concerned, believing that ChatGPT lacks the true intelligence to understand the complexities of human language and conversation (Bogost 2022). While the limitations of the generative AI are evident, such as the potential to generate inaccurate information, produce harmful instructions, present biased content, and have limited knowledge due to the data on which it was trained, many educators perceive this as an opportune moment to revolutionize teaching and learning amidst the challenges presented by ChatGPT.[3]

---

[1] https://www.celt.iastate.edu/resources/ai-teach-learn/
[2] https://www.sportskeeda.com/gaming-tech/all-exams-chatgpt-passed-far

In order to address the challenges and opportunities posed by generative AI tools such as ChatGPT, it is important for universities to have policies in place to ensure that ChatGPT is used ethically and responsibly in teaching and learning. The university's policy on generative AI tools serves as a catalyst for institutional and pedagogical change. A comprehensive yet flexible policy could enable faculty and students to reap the benefits of using such technology in the classroom. The goals of a ChatGPT policy generally include preventing cheating by establishing clear guidelines on whether, when and how the tools can be used, raising student awareness of the potential risks and benefits of using ChatGPT (e.g., data privacy), and ensuring that ChatGPT is used to enhance student learning. While some universities have swiftly implemented new policies in response to ChatGPT, the majority are taking a wait-and-see approach, monitoring developments before taking action. According to Veletsianos, Kimmons, and Nondah (2023), there is limited literature on how universities are actively engaging with and responding to generative AI tools in the classrooms.

This study aims to examine university policies regarding the use of ChatGPT in teaching and learning. A university ChatGPT policy is defined as a set of guidelines and procedures that govern the actions and decisions of a university or a college regarding generative AI tools such as ChatGPT. Specifically, we examine the following two research questions: (1) What is the status and trend of university ChatGPT policies? and (2) What are the factors that are associated with universities' decisions to implement different ChatGPT policies (i.e., whether to issue a university-wide policy or not, and ban vs. embrace ChatGPT)? To answer these questions, we collected information on ChatGPT policies issued in universities worldwide. We focus on the top 500 universities according to the 2022 Quacquarelli Symonds (QS) World University Rankings.[4] We first identified the existence of a ChatGPT policy in these 500 universities; then we downloaded the university's ChatGPT policies and analyzed their content, including whether ChatGPT is embraced or banned in universities and the extent to which the policy covers on ChatGPT usage. We also collected the issuance dates of ChatGPT policies.

---

[3] https://teaching.missouri.edu/blog/chatgpt-generative-ai

We observed significant heterogeneity in the issuance of ChatGPT policies among the top 500 universities. Notably, less than a third (26.53%) of the top 500 universities have responded with a ChatGPT. Of those universities with a ChatGPT policy, approximately 70.4% are located in the countries where English is the native language, including the United States, United Kingdom, Canada, and Australia. It is also worth noting that there are different patterns of policy issuance among universities in different countries and regions. Some regions exhibit a high percentage (over 50%) of universities with a ChatGPT policy, whereas others have yet to implement such a policy. This shows a dispersed geographical distribution of responses to the challenges and opportunities presented by generative AI tools.

Of the universities that have implemented a policy, a total of 43 universities (32.6%) have chosen to ban ChatGPT by restricting the use of ChatGPT or any other AI tools in assessments unless explicitly permitted. Among these universities, 14 strictly prohibit students from using AI chatbots like ChatGPT during exams, even in exams that allow the use of exam aids. The other universities prohibit the use of ChatGPT in assessments but give individual instructors the discretion to deviate from this strict policy. However, any deviation from the university's ChatGPT policy must be consistent with the institution's academic integrity rules. The percentage of universities that have a policy to ban ChatGPT varies across countries, ranging from 0% to 100%. On average, the mean percentage is 29.9%, with a standard deviation of 35.9%.

---

[4] https://www.topuniversities.com/university-rankings/world-university-rankings/2022

When analyzing the temporal pattern of university ChatGPT policy issuance, it is noteworthy that universities in Denmark, the United States, and the United Kingdom were among the early responders, issuing their policies within two months of ChatGPT's launch. This observation suggests that European and U.S. universities may be nimbler when it comes to providing policy governing the use of a new technology in education. The time it takes for universities to implement a ChatGPT policy varies significantly, ranging from 34 days to 153 days, highlighting the considerable variation among institutions. In addition, the number of universities that have policies embracing ChatGPT is outpacing those that have chosen to prohibit its use, indicating a growing acceptance of the tool within the academic community.

In examining the scope of current university ChatGPT policies, it is clear that the primary focus is on establishing academic integrity policies to address the use of ChatGPT in assessments including assignments, exams, and essay writing. Universities have taken steps to revise their existing academic integrity or misconduct policies to clearly outline the specific forms of ChatGPT use that would be considered plagiarism or cheating. Of the universities with ChatGPT policies (either ban or embrace), 19 universities (approximately 14%) have provided students with guidelines on how to use ChatGPT in an appropriate and ethical manner. These guidelines include instructions on how to properly cite and reference the use of AI or ChatGPT in assessment work, as well as reminders of the potential risks and benefits associated with its use. Another 40 universities (30%) have provided teaching guidelines for instructors, which include sample syllabus statements that specify whether and to what extent students may incorporate ChatGPT into their coursework.

To explore the factors that are associated with universities' ChatGPT policy decisions, we are particularly interested in examining the following factors: (1) university characteristics, such as the university's reputation and the geographic region in which the university is located; (2) the ChatGPT policies of peer universities in the same country; (3) the ChatGPT sentiment measured by Google trend metrics of the keyword search for "ChatGPT," which indicates the general public's knowledge of and attention to ChatGPT.

We conduct a logit model analysis to examine how these factors are influencing universities' ChatGPT policies. The logit model analysis shows that the following factors are significantly and positively associated with university's likelihood of having a ChatGPT policy: a university's academic reputation score, whether the university is located in one of the four English-speaking countries with English as the native language, and the Google trend metrics of the keyword search for "ChatGPT". In addition, we found that the likelihood of a university having a policy banning ChatGPT is positively associated with the faculty student ratio, the citations, and negatively associated with the number of peer universities in the same country that have banned ChatGPT.

Our study provides valuable academic and practical insights. The findings help policymakers understand how universities are adopting emerging AI technologies and make informed intervention decisions when necessary. In addition, our findings help institutions understand the current landscape of ChatGPT in higher education sector, and how their peer universities are developing policies to guide faculty and students in the use of these new technology tools and facilitate their decision-making.

1.1 Literature

Our study is closely related to emerging studies which study the ChatGPT-related issues in the education sector. A recent study by Eke (2023) examines the ethical issues, particularly academic integrity, raised by allowing students to outsource their writing to ChatGPT. Sullivan et al. (2023) conducted a content analysis of 100 news articles related to university responses and public discussions about ChatGPT in Australia, New Zealand, the United States, and the United Kingdom. They found that public discussions and university responses focused primarily on concerns about academic integrity and opportunities for innovative assessment design, with very little discussion of student perspectives. Neither study analyzes the content of university ChatGPT policies or examines the factors that influence university ChatGPT policies. Our work contributes to this literature by examining the policy making of generative AI in higher education markets. We study universities' decisions on whether to have a ChatGPT policy and whether to ban ChatGPT in classrooms and discuss the implications for product design in higher education.

Our work is also related to a rapidly growing and dynamic literature that explores policy making and policy regulation related to generative AI. For example, Jo (2023) highlights the importance of policy regulations that enforce the honest use of large language models, such as ChatGPT. Sarel (2023) focuses on the various proposals to restrict AI in the European Union, arguing that these proposals may distort incentives and lead to inefficiencies. Hacker, Engel, and Mauer (2023) situate the emerging generative models in the current debate on AI regulation and suggest that more differentiated regulation would better fit the realities of the evolving AI value chain. de Rassenfosse, Jaffe, and Wasserman (2023) propose that current policy support and regulations for patents need to be adapted to accommodate AI-generated inventions. In this paper, we focus on the policy regulations pertaining to generative AI in higher education. Specifically, we examine ChatGPT policies implemented at universities worldwide, including their existence, content (embracing or banning ChatGPT, and the coverage of policy), and issuance date (i.e., how long it will take for a university to respond).

Our paper is also generally related to the literature that studies the education sector, which plays a critical role in cultivating and shaping the future of individuals and societies. Beginning with Becker (1964), economists have explored it through the lens of human capital and its impact on productivity. A more recent stream of economic literature investigates the role of policy regulations in promoting equitable access to quality education and improving student learning outcomes. Chan and Eyster (2003) study the effects of banning a policy that grants advantages to disadvantaged students in college admissions and show that such a ban always reduces diversity and could also reduce student quality. Fu (2006) finds that a minority-favored admissions policy could widen the racial test score gap. Krishna and Tarasov (2016) show that an admissions policy that favors the disadvantaged group may be desirable when education is costly and easy to obtain, and undesirable otherwise. Moreover, there is an emerging and growing body of marketing research on education and policy regulation (Bao, Ni, & Singh, 2023; Bao, Jerath, & Singh, 2023).

## 2. Institutional Background and Data

We gather information about ChatGPT policies at universities around the world, including determining whether a policy has been issued, analyzing the content of the policy, and identifying the date when the policy was issued. When analyzing the content of a university's ChatGPT policy, we focus on the following aspects: whether ChatGPT is embraced or banned, and for universities with ChatGPT policies, we analyze the details of the policies, such as whether they provide guidance to faculty or student on ChatGPT and whether the university updates its academic integrity code to govern the use of ChatGPT. As defined earlier, a university ChatGPT policy refers to the policies and procedures that govern a university's action and decisions regarding generative AI tools such as ChatGPT. We focus on the top 500 universities according to the 2022 Quacquarelli Symonds (QS) World University Rankings. We identify the university's ChatGPT policy by searching each university's official website.

Official university websites typically present various information about the institution, its programs, and its services and communicate the important information to the university community. A university is considered to have a ChatGPT policy if we can find the policy clearly stated on its official website by the end of our sample period (May 2$^{nd}$, 2023).[5] For example, the University of Cambridge issued its ChatGPT policy by revising its Plagiarism and Academic Misconduct policy on April 2023, stating that "The University has strict guidelines on student conduct and academic integrity. These stress that students must be the authors of their own work. Content produced by AI platforms, such as ChatGPT, does not represent the student's own original work so would be considered a form of academic misconduct to be dealt with under the University's disciplinary procedures."[6]

When analyzing the content of ChatGPT policies, we identified two primary types of policies. The first type is to embrace ChatGPT, i.e., ChatGPT can be freely used in accordance with academic integrity conduct or there are no university-level specific restrictions on the use of ChatGPT. The second type is a prohibition of using ChatGPT, i.e., ChatGPT is prohibited in at least one scenario (activity) or more scenarios on the campus.

When collecting information on whether a university has issued a ChatGPT policy, we also noted the date when the policy was issued. Once the university issues a policy in response to the generative AI tool, the students and faculty will have clear guidelines for using ChatGPT. Issuing the policy early will minimize confusion and anxiety and benefit the university community. Using this data, we calculated "*ChatGPT policy day*" as the number of days that have passed since the introduction of ChatGPT. Note that this measurement includes the day when ChatGPT was introduced.

---

[5] We acknowledge the incidence that a university may have issued a ChatGPT policy but not made it publicly available on their official website. As a result, our sample may not be able to accurately identify such cases.
[6] https://www.plagiarism.admin.cam.ac.uk/what-academic-misconduct/artificial-intelligence.

According to the QS World University Rankings, the top 500 universities are distributed across 58 countries and regions. Table A1 in the online appendix provides a summary of this distribution. The United States has 87 universities in the top 500, the highest number among countries and regions, followed by the United Kingdom with 49 universities, Germany with 31 universities, and Australia and China with 26 universities each. The remaining 282 universities are spread across 53 countries and regions. Of the top 500 universities, we found that 132 universities (26.35%) in 22 countries and regions have issued their ChatGPT policies, with 43 (32.6%) in the United States, 23 (17.4%) in the United Kingdom, 18 (13.6%) in Australia, and 9 (6.8%) in Canada. Essentially, the issuance of ChatGPT policies among universities is heterogenous, and the four English-speaking countries with English as the native language represent 70.4% of the universities with ChatGPT policies.

Policy issuance patterns also vary across countries and regions. Among countries with more than one university in the QS World University Rankings 2022, the majority of universities in Hong Kong SAR have issued ChatGPT policies, with 5 out of 6 (83.33%) having done so. In Australia, 18 out of 26 universities (69.23%) have issued such policies, followed by 3 out of 5 universities (60%) in Denmark, and 9 out of 17 universities (52.94%) in Canada. However, in Germany, Japan and France, less than 10% of universities have implemented ChatGPT policies. Table 1A summarizes this pattern of data. The absence of universities from certain countries listed in the top 500 implies that none of their universities have implemented ChatGPT policies.

We also examined the relationship between universities' decision to implement a ChatGPT policy and their rankings in the QS World University Rankings. We divided the top 500 universities into five rank bands, with those in the top 100 in band 1, and those between 101st and 200th in band 2, and so on. Table 1B shows that universities with ChatGPT policies are predominantly found in rank bands 1 and 2, indicating that higher-ranked universities are more likely to have adopted this generative tool. In our logit model analysis, we examine how the likelihood of a university having a ChatGPT policy is associated with the reputation of the university.

==== INSERT TABLES 1A AND 1B AND FIGURE 1 ABOUT HERE ====

Of the 132 top 500 universities with a ChatGPT policy, 89 (67.43%) embraced this new generative AI tool, more than double the number of universities that banned it. Table 2A summarizes the distribution of these universities' policies by type and country. The ratio, defined as the number of universities in a country that have a policy banning ChatGPT over the number of universities that have a ChatGPT policy, varies from 0% to 100%, with a mean of 29.9% and a standard deviation of 35.9%.

We further examined how the distribution of these universities' policies differed by rank band. Table 2B summarizes the distribution of these universities' policies by type and rank. The highest ratio of universities with a policy banning ChatGPT to universities with a ChatGPT policy is observed in rank band 3, followed by rank band 1. In our model analysis, we will explore how the likelihood of a university issued a policy that banned ChatGPT is associated with a university's reputation. ChatGPT has been banned at the country-level in countries in our sample such as China, Iran, and Russia.[7][8] Universities located in these countries that rank among the top 500 in the QS World University Rankings are considered to have no ChatGPT policy yet. In our empirical model analysis, we conducted a robustness analysis that excluded these universities from the sample altogether.

---

[7] https://www.digitaltrends.com/computing/these-countries-chatgpt-banned/.
[8] On March 31, 2023, Italy's data regulator Garante temporarily banned ChatGPT over data security concerns. Access to the ChatGPT chatbot has been restored in Italy on April 28, 2023.

==== INSERT TABLES 2A AND 2B AND FIGURE 2 ABOUT HERE ====

We are able to identify the policy date information for 126 (95.4%) out of the 132 universities. As shown in Table 3A, none of the universities responded with a university policy in the first month after the introduction of ChatGPT, 9 universities introduced the ChatGPT policy in the second month, followed by 28 universities in the third month, 44 and 41 in each of the following months, and 4 in the two days of last month of our sample period. Note that the cutoff date of our sample is May 2$^{nd}$, so there are only two days in the last month (i.e., May 2023) of our sample period. Figure 1A shows how the cumulative number of universities with ChatGPT policies and the number of universities issuing their ChatGPT policies with the number of days elapsed since the introduction of ChatGPT. Figure 1B illustrates the change in the cumulative number of universities embracing and banning ChatGPT over time. The number of universities embracing ChatGPT is increasing faster than the number of universities banning it. We also examined the distribution of the number of universities that implemented a ChatGPT policy each month, broken down by region in Table 3B. Universities from Denmark, the United States and the United Kingdom were among the first group to respond to ChatGPT, issuing their policies in the second month after its introduction. Following their lead, universities in Australia, Canada, Finland, Hong Kong SAR, and the Netherlands have also implemented their own ChatGPT policies.

==== INSERT TABLES 3A AND 3B ABOUT HERE ====

==== INSERT FIGURES 1A AND 1B ABOUT HERE ====

Based on our data sample, we found that Aarhus University from Denmark was the first university in the top 500 universities to implement a ChatGPT policy. They did so on January 3rd, 2023, approximately 34 days after ChatGPT became available to the public. Aarhus University takes a strict stance on the use of ChatGPT policy, as outlined in their guidelines for students, they explicitly state that the use of ChatGPT is not allowed in exams, even if it is allowed by the instructor. Seventeen days later, Harvard University issued its ChatGPT policy but took a relatively more flexible approach. In its guidance to faculty, Harvard University informs them that students are generally prohibited from using ChatGPT or other AI tools in assessed work; however, they allow colleges or individual instructors to deviate from this policy. Although Harvard University also imposes restrictions on ChatGPT, their policy differs from Aarhus University's in the way that Harvard allows individual instructors the discretion to deviate from the university-wide policy, while Aarhus does not allow such deviations. New York University (NYU) was the first to announce a university-wide policy that embraces ChatGPT in teaching. On January 21st, 2023, they explicitly state that there are essentially no restrictions on students' use of AI tools in courses at the university level and that faculty members are expected to inform students about the extent to which AI tools, including ChatGPT, can be used in their coursework. Thus, NYU's policy provides considerable latitude for different programs and disciplines to decide how to integrate ChatGPT into their teaching practices.

The issuances of ChatGPT policies varied in terms of the number of days it took, ranging from 34 days to 153 days. The average was 107 days, with a median of 113 days. To provide further insight, we analyzed this distribution by country and present the results in Table 4A. It is evident that the time taken to impose a ChatGPT policy varies across countries. Among countries with multiple universities have issued ChatGPT policy on known date, the Netherland had the fastest average response time of 87.8 days, followed by Denmark (90.6 days) and the United States (98.8 days).  This suggests that universities in Europe and North America are more responsive to the need for policy support for new technology-enabled teaching and learning tools. We also observed a non-linear association between the time taken to adopt ChatGPT policies and university rank.Universities in the middle rank band showed a faster response than those in higher or lower rank bands.The empirical distribution of *ChatGPT policy day*, categorized by rank band, can be found in Table 4B.

==== INSERT TABLES 4A AND 4B ABOUT HERE ====

We also examined the extent to which universities' ChatGPT policies detail the legitimate uses of ChatGPT. In general, there are three ways to address the use of generative AI tools such as ChatGPT can be reflected in the university's ChatGPT policies.  One approach is to augment existing policies, such as academic integrity rules, honor codes, or student misconduct policies, by clearly specifying the types of ChatGPT use that would violate those rules and constitute plagiarism and cheating. While this approach provides a broad framework, it lacks the flexibility to tailor ChatGPT policies to specific programs or disciplines. 34 out of 132 universities with ChatGPT policies have amended their existing academic integrity policies to explicitly state that inappropriate use of AI tools such as ChatGPT is a violation of academic integrity.

The second approach is to provide student guidelines that specify the conditions under which students may use ChatGPT in assessments and coursework. In total, 19 universities in our sample have student guidelines, but the extent to which the student guidelines are comprehensive varies widely. Some guidelines provide specific scenarios for the use of AI. For example, several universities do not prohibit the use of ChatGPT, but require that the use of AI tools in coursework should be transparent and acknowledged. While some have provided examples or explanations of how AI was used to generate ideas, draft content, or develop structures. Even the prompts used to generate the content should be included in the statements. The others provide guidance on how to cite and reference AI tools in the text. Another typical content included in the guidelines for students is highlighting the limitations of generative AI and the dangers of relying on it.

The third approach is to provide instructors with guidelines for using ChatGPT in the classroom. Of the 133 campuses with ChatGPT policies, 40 campuses have guidelines for instructors covering a range of topics, including syllabus design, AI statement in syllabi, assessment design, classroom use of ChatGPT, and setting rules for students and reminding students that the Academic Integrity Policy applies to their coursework.

3. **Empirical Analysis**

We start with examining how different factors are associated with universities' decisions to issue or not issue a ChatGPT policy on day *t*. Specifically, whether a university issued its ChatGPT policy on day *t* in our sample period (a binary choice) was modelled as:

$$y_{it} = \begin{cases} 1 \text{ if } y_{it}^* > 0 \\ 0 \text{ otherwise} \end{cases}$$

where $y_{it}^* = \beta_0 +$

$$\underbrace{\beta_1 \cdot University\ reputation_i + \beta_2 \cdot English\ speaking_i}_{\text{university characteristics}} + \underbrace{\beta_3 \cdot number\ of\ universities\ issuing\ policy_{it}}_{\text{peer effects}} + \underbrace{\beta_5 \cdot ChatGPT\ sentiment_i}_{\text{external factors}} + C'_{it}\alpha + u_{it}$$ (1)

where $y_{it}^*$ is a latent continuous variable representing university *i*'s propensity to issue a ChatGPT policy on day *t*. The policy variable $y_{it}$ takes the value 1 if $y_{it}^*$ exceeds the threshold of 0, otherwise it takes the value 0. Each university has $t = 1, \ldots, \tau_i$ observations, where $\tau_i$ represents the *ChatGPT policy day*, the day that university *i* issued its policy. For universities that have not yet issued a ChatGPT policy in our sample period, $\tau_i$ is the length of the data sample period (i.e., $\tau_i = 153$).

We include the following explanatory variables. *University reputation* is a vector of variables measuring the reputation of university. The variable *university rank* is the rank of a university in the QS World University Rankings. A lower value of this variable indicates that a university is ranked higher and has a higher overall reputation. In reporting a university's rank, QS World University Rankings also provides a university's score in each of the following six indicators: *academic reputation*, *employer reputation*, *faculty/student ratio*, *citations per faculty*, *international student ratio*, and *international faculty ratio*. We also included a university's score in these six indicators as a part of the *university reputation* vector. *English speaking* is a university characteristic variable that indicates whether the university is located in one of the following five English-speaking countries where English is the native language: the United Kingdom, the United States, Canada, Australia, New Zealand. We measure the peer effect using the *number of universities issuing policy*, which is the number of peer universities in the same country that issued the ChatGPT policy before the focal university. To reduce the skewness, we applied a logarithmic transformation to this variable. *ChatGPT sentiment* is measured by Google trend metrics of the keyword searches for "ChatGPT". We collected Google trend metrics of the keyword search for "ChatGPT" from November 1, 2022 to May 2, 2023 and used it as an indicator of the public interest in ChatGPT. $C'_{it}$ includes the control variables including country fixed effects. $u_{it}$ denotes the error term, which follows a Type-1 extreme value distribution.

We then investigated the factors associated with a university's decision to ban ChatGPT conditioning on issuing a policy. To conduct this analysis, we used a dependent variable denoted $d_i^*$, which is a latent continuous variable indicating a university's propensity to issue a policy banning ChatGPT. The decision variable $d_i$ takes a value of 1 if $d_i^*$ exceeds the threshold of 0 and 0 otherwise. The explanatory and control variables used in the analysis remained the same except that we replaced the peer effect variable with the number of peer universities within the same country that had already banned ChatGPT before the focal university.

Results

Table 5 shows the summary statistics and correlation coefficients for the variables used in this study. None of the correlation coefficients have a high value, indicating that there is multicollinearity problem.

==== INSERT TABLE 5 ABOUT HERE ====

Table 6 reports the results of the university's decision to issue a ChatGPT policy during the sample period. Among the variables in the vector of university reputation, the coefficient of *university rank* is negative but not statistically significant ($\beta = -0.001$, $p = 0.491$). Among the six indicators, the coefficient of academic reputation is positive and significantly significant ($\beta = 0.018$, $p = 0.032$), indicating that the likelihood of a university issuing its ChatGPT policy is positively associated with its academic reputation score in theQS World University Rankings. This implies that a university with a higher academic reputable is more likely to issue a ChatGPT policy. The effects of other indicators such as *employer reputation*, *faculty/student ratio*, citation, *international student ratio,* and *international faculty ratio* are insignificant.

In addition, whether the university is located in one of the major English-speaking countries is positively associated with the likelihood of a university issuing a ChatGPT policy ($\beta = 1.057$, $p = 0.044$). Regarding the peer effect, the coefficient of the *number of universities issuing policy* is negative but not statistically significant ($\beta = -0.129$, $p = 0.390$). Finally, ChatGPT sentiment is significant and positively associated with the likelihood of a university issuing ChatGPT policies ($\beta = 0.045$, $p = 0.000$), suggesting that a university's chance to issue a ChatGPT policy is positively associated with how the ChatGPT issue becomes more prominent and important in the social media.

Table 7 reports the results of the university's decision to issue a policy to ban ChatGPT in a day during the sample period. Regarding the variables in the *university reputation* vector, the coefficient of *university rank* is positive but statistically insignificant ($\beta = 0.006$, $p = 0.243$). The sign indicates that the likelihood of a university to issue a policy to ban ChatGPT is positively associated with its rank in the QS World University Rankings, and a lower ranked university is more likely to have a policy to ban ChatGPT. Among the six indicators, the coefficient of faculty student ratio is positive and statistically significant ($\beta = 0.031$, $p = 0.021$), indicating that the likelihood of a university to issue a policy to ban ChatGPT is positively associated with faculty-student ratio. This shows that a university with a higher faculty-student ratio may have a higher chance of issuing a policy to ban ChatGPT. The coefficient of citation is also positive and statistically significant ($\beta = 0.033$, $p = 0.054$), indicating that a university's probability of issuing a policy to ban ChatGPT is positively associated with citations. This shows that a university with more citations may have a higher chance of having a policy to ban ChatGPT. In addition, the coefficient of English speaking is positive and statistically significant ($\beta = 2.985$, $p = 0.063$), indicating that a university in the English-speaking countries with English as the native language is more likely to ban ChatGPT, conditional on having a ChatGPT policy.

Regarding the peer effect, the coefficient of the number of peer universities in the same country that had banned ChatGPT is negative and statistically significant ($\beta = -1.838$, $p = 0.013$), indicating that the likelihood of a university issuing a policy to ban ChatGPT is negatively associated with the number of peer universities in the same country that had banned ChatGPT.

==== INSERT TABLES 6 AND 7 ABOUT HERE ====

## 4. Discussions and Conclusion

In this study, we provide a comprehensive overview of the current landscape regarding universities implementing ChatGPT policies on campuses. In addition, we present preliminary empirical findings that shed light on the factors influencing universities' varied responses to ChatGPT.

Our study reveals substantial variation in the adoption of ChatGPT policies among the top 500 universities in the QS World University Rankings, with less than one-third of the universities have implemented ChatGPT policies. Among the universities that implemented ChatGPT policies, an overwhelming majority of 67.4% chose to embrace ChatGPT, more than doubling the number of universities that chose to prohibit its use in coursework. Notably, among the universities that chose to ban ChatGPT, there was a noticeable difference in the degree of flexibility given to faculty to deviate from university policy, although a majority of these universities allow some degree of flexibility. These observations suggest that a significant number of universities have acknowledged the importance of tailored policies that recognize the diverse applications of ChatGPT in higher education. Various majors or disciplines may find generative AI to be a useful tool. Thus, upholding a flexible policy framework would enable universities and academic programs to delve into how ChatGPT can revolutionize teaching and learning within their respective fields while addressing the associated challenges for its implementation.

Our findings reveal distinct patterns in policy issuance across countries and regions, particularly in terms of the time taken by universities to issue ChatGPT policies. European and North American universities show a higher level of responsiveness to new technology tools, as evidenced by their more timely provision of policy support. This observation underscores the geographical dispersion in the distribution of knowledge about generative AI.

Our empirical model analysis reveals that several factors, including the academic reputation score, being located in an English-speaking country, and sentiment toward "ChatGPT," are positively and statistically significant in relation to the probability of a university implementing a ChatGPT policy. We also found that the likelihood of a university issuing a policy banning the use of ChatGPT is positively associated with its faculty-student ratio, citations, and location (i.e., English-speaking country), while negatively associated with the number of peer universities in the same country that have banned ChatGPT.

Universities that have chosen to ban ChatGPT face the challenge of effectively detecting the use of ChatGPT. As generative AI tools continue to advance and provide more robust alternatives to ChatGPT, it becomes increasing difficult to prevent students from accessing and using them. Despite ongoing efforts to develop tools to detect the use of ChatGPT in student work, the reliability of such tools remains questionable. Given these challenges, Florida State University emphasizes the importance of universities working with students to ensure that they have a full understanding of the proper use of these tools.

In reviewing the university's policies, it is clear that concerns about cheating, plagarithm, and other activities that violate academic integrity are driving the decision to ban ChatGPT. However, it is interesting to note that some universities have chosen to embrace ChatGPT in teaching while implementing policies to prevent academic integrity violations. This underscores the need for a comprehensive approach that carefully considers the potential risks and safeguards associated with integrating generative AI tools into academia. As indicated in the website of RMIT, the question that need more explorations by educators and policy makers are how we can take good use of ChatGPT for better teaching and learning. Our challenge as instructors is to learn how we can use new tools authentically and creatively and to support creativity while maintaining academic integrity. Essentially, how will instructors design AI-assisted learning and assessment?

In developing a comprehensive policy for the use of ChatGPT in education, several key considerations should be taken into account. First, in addition to adapting academic integrity policies to govern the use of generative AI, universities need to provide clear guidelines for students to help them understand the appropriate use of ChatGPT and its limitations. Second, the policy should emphasize the importance of ethical and moral use of ChatGPT. When students are allowed to use generative AI tools such as ChatGPT, it becomes important to determine the breaking point at which an essay can no longer be attributed solely to the student's efforts. As a result, universities should provide guidelines or training on the responsible use of AI, including proper citation and referencing when using AI-generated content. Third, universities should provide a training program for faculty to ensure that they have the necessary knowledge and skills to design courses and assessments that integrate ChatGPT into their teaching. In addition, the university policy should establish mechanisms for regularly evaluating the effectiveness and impact of ChatGPT on students' learning, thereby identifying potential problems and challenges for future improvement. Fourth, the policy should encourage collaboration between faculty, students, and other stakeholders (e.g., AI tool developers) in the policy-making process, which will help shape and refine the guidelines for ChatGPT use. Finally, the policy should include a degree of flexibility to accommodate the evolving nature of generative AI technology.

As a first study exploring universities' policy issuance decisions for generative AI tools such as ChatGPT, our research has certain limitations. First, during the data collection process, if we did not find a specific policy, we inferred that the university did not have a ChatGPT policy during the sample period. However, the absence of a policy does not indicate whether the university is actively planning the policy or has not yet begun. Distinguishing between these two scenarios is challenging, and our observations are primarily focused on whether a ChatGPT policy has been officially issued by a university. In cases where no policy has been explicitly issued, we are unable to determine the underlying status of the decision-making process. Differentiating between these scenarios could provide additional insights. Second, our sample is limited to data collected until May 2nd, 2023. Any developments or changes that occurred after that date are not accounted for in our study. The future research aims to gather additional data to integrate the latest policy advancements in the adoption of generative AI tools within universities.

Table 1A: Summary of # of University Having Policy for Generative AI Tool (e.g., ChatGPT) by Country (Region)

| Country (Region) | # of Top 500 Universities | # of Universities having ChatGPT policy | The rate of Universities having ChatGPT policy (%) |
|---|---|---|---|
| United States | 87 | 43 | 49.43 |
| United Kingdom | 49 | 23 | 46.94 |
| Germany | 31 | 2 | 6.45 |
| Australia | 26 | 18 | 69.23 |
| Canada | 17 | 9 | 52.94 |
| Japan | 16 | 7 | 43.75 |
| South Korea | 16 | 1 | 6.25 |
| Netherlands | 13 | 5 | 38.46 |
| France | 11 | 1 | 9.09 |
| Taiwan | 10 | 1 | 10.00 |
| Switzerland | 9 | 2 | 22.22 |
| Belgium | 8 | 1 | 12.50 |
| New Zealand | 8 | 3 | 37.50 |
| Sweden | 8 | 2 | 25.00 |
| Finland | 7 | 1 | 14.29 |
| Hong Kong SAR | 6 | 5 | 83.33 |
| Denmark | 5 | 3 | 60.00 |
| Ireland | 5 | 1 | 20.00 |
| South Africa | 4 | 1 | 25.00 |
| Chile | 3 | 1 | 33.33 |
| Estonia | 1 | 1 | 100.00 |
| Macau SAR | 1 | 1 | 100.00 |

Table 1B: Summary of # of University Having ChatGPT Policy by Rank Band

| Rank | # of universities having ChatGPT policy | Percent (%) |
|---|---|---|
| 1-100 | 50 | 38 |
| 101-200 | 34 | 26 |
| 201-300 | 18 | 14 |
| 301-400 | 17 | 13 |
| 401-500 | 13 | 10 |

**Table 2A**: Summary of # of Universities Having ChatGPT Policy by Type of Decision and Region

| Country (Region) | # of universities having policy to ban ChatGPT | Percent of universities having policy to ban ChatGPT | # of universities having policy to embrace ChatGPT | Percent of universities having policy to embrace ChatGPT |
|---|---|---|---|---|
| Australia | 5 | 27.8 | 13 | 72.2 |
| Belgium | 0 | 0 | 1 | 100 |
| Canada | 2 | 22.2 | 7 | 77.8 |
| Chile | 0 | 0 | 1 | 100 |
| Denmark | 3 | 100 | 0 | 0 |
| Estonia | 0 | 0 | 1 | 100 |
| Finland | 0 | 0 | 1 | 100 |
| France | 1 | 100 | 0 | 0 |
| Germany | 2 | 100 | 0 | 0 |
| Hong Kong SAR | 4 | 80 | 1 | 20 |
| Ireland | 0 | 0 | 1 | 100 |
| Japan | 3 | 42.9 | 4 | 57.1 |
| Macau SAR | 0 | 0 | 1 | 100 |
| Netherlands | 2 | 40 | 3 | 60 |
| New Zealand | 1 | 33.3 | 2 | 66.7 |
| South Africa | 0 | 0 | 1 | 100 |
| South Korea | 0 | 0 | 1 | 100 |
| Sweden | 1 | 50 | 1 | 50 |
| Switzerland | 0 | 0 | 2 | 100 |
| Taiwan | 0 | 0 | 1 | 100 |
| United Kingdom | 9 | 39.1 | 14 | 60.9 |
| United States | 10 | 23.3 | 33 | 76.7 |

**Table 2B**: Summary of # of University Having ChatGPT Policy by Type of Decision and Rank Band

| rank_band | # of universities having policy to ban ChatGPT | Percent of universities having policy to ban ChatGPT | # of universities having policy to embrace ChatGPT | Percent of universities having policy to embrace ChatGPT |
|---|---|---|---|---|
| 1 | 20 | 40 | 30 | 60 |
| 2 | 10 | 29.4 | 24 | 70.6 |
| 3 | 8 | 44.4 | 10 | 55.6 |
| 4 | 4 | 23.5 | 13 | 76.5 |
| 5 | 1 | 7.7 | 12 | 92.3 |

**Table 3A**: # of Universities Introducing ChatGPT Policy by Months

| Month ID | Year/Month | # of universities | Percentage |
|---|---|---|---|
| 1 | 2022/12 | 0 | 0 |
| 2 | 2023/01 | 9 | 7.1 |
| 3 | 2023/02 | 28 | 22.2 |
| 4 | 2023/03 | 44 | 34.9 |
| 5 | 2023/04 | 41 | 32.5 |
| 6 | 2023/05 | 4 | 3.2 |

Note: in the last month of our sample period, May, there are only two days included, namely May 1st and May 2nd.

**Table 3B**: # of Universities Introducing ChatGPT Policy by Month and Country

| Country (Region) | First month | Second month | Third month | Fourth month | Fifth month | Sixth month |
|---|---|---|---|---|---|---|
| Australia | 0 | 0 | 5 | 6 | 6 | 1 |
| Belgium | 0 | 0 | 0 | 0 | 0 | 0 |
| Canada | 0 | 0 | 2 | 4 | 2 | 0 |
| Chile | 0 | 0 | 0 | 0 | 1 | 0 |
| Denmark | 0 | 1 | 0 | 1 | 1 | 0 |
| Estonia | 0 | 0 | 0 | 0 | 1 | 0 |
| Finland | 0 | 0 | 1 | 0 | 0 | 0 |
| France | 0 | 0 | 0 | 1 | 0 | 0 |
| Germany | 0 | 0 | 0 | 2 | 0 | 0 |
| Hong Kong SAR | 0 | 0 | 1 | 2 | 2 | 0 |
| Ireland | 0 | 0 | 0 | 1 | 0 | 0 |
| Japan | 0 | 0 | 0 | 0 | 7 | 0 |
| Macau SAR | 0 | 0 | 0 | 0 | 1 | 0 |
| Netherlands | 0 | 0 | 4 | 0 | 1 | 0 |
| New Zealand | 0 | 0 | 0 | 2 | 0 | 1 |
| South Africa | 0 | 0 | 0 | 1 | 0 | 0 |
| South Korea | 0 | 0 | 0 | 1 | 0 | 0 |
| Sweden | 0 | 0 | 0 | 1 | 1 | 0 |
| Switzerland | 0 | 0 | 0 | 0 | 2 | 0 |
| Taiwan | 0 | 0 | 0 | 1 | 0 | 0 |
| United Kingdom | 0 | 1 | 3 | 11 | 5 | 1 |
| United States | 0 | 7 | 12 | 10 | 11 | 1 |

Note: ChatGPT policy day is missing for the university which has issued policy in Belgium.

**Table 4A**: Empirical Distribution of ChatGPT Policy Day by Country

| Country (Region) | Number | Mean | Median | Standard deviation | Min | Max |
|---|---|---|---|---|---|---|
| Australia | 18 | 111.8 | 113 | 30.6 | 64 | 153 |
| Belgium | 1 | | | | | |
| Canada | 9 | 107.1 | 102 | 23.0 | 85 | 144 |
| Chile | 1 | 122 | 122 | | 122 | 122 |
| Denmark | 3 | 90.7 | 94 | 55.1 | 34 | 144 |
| Estonia | 1 | 149 | 149 | | 149 | 149 |
| Finland | 1 | 78 | 78 | | 78 | 78 |
| France | 1 | 113 | 113 | | 113 | 113 |
| Germany | 2 | 115.0 | 115 | 5.7 | 111 | 119 |
| Hong Kong SAR | 5 | 107.8 | 118 | 24.5 | 65 | 124 |
| Ireland | 1 | 110 | 110 | | 110 | 110 |
| Japan | 7 | 133.6 | 138 | 10.2 | 122 | 148 |
| Macau SAR | 1 | 132 | 132 | | 132 | 132 |
| Netherlands | 5 | 87.8 | 77 | 31.4 | 63 | 142 |
| New Zealand | 3 | 120.3 | 113 | 28.7 | 96 | 152 |
| South Africa | 1 | 113 | 113 | | 113 | 113 |
| South Korea | 1 | 106 | 106 | | 106 | 106 |
| Sweden | 2 | 115.5 | 115.5 | 33.2 | 92 | 139 |
| Switzerland | 2 | 122 | 122 | 0 | 122 | 122 |
| Taiwan | 1 | 91 | 91 | | 91 | 91 |
| United Kingdom | 23 | 110.7 | 113 | 29.7 | 54 | 152 |
| United States | 43 | 98.8 | 91 | 32.0 | 51 | 152 |

Note: ChatGPT policy day is missing for the university which has issued policy in Belgium.

**Table 4B**: Empirical Distribution of ChatGPT Policy Day by Rank

| Rank band | Number | Mean | Median | Standard deviation | Min | Max |
|---|---|---|---|---|---|---|
| 1 | 50 | 107.1 | 115 | 29.4 | 51 | 153 |
| 2 | 34 | 102.4 | 110 | 29.7 | 34 | 152 |
| 3 | 18 | 97.5 | 104 | 28.9 | 54 | 152 |
| 4 | 17 | 111.2 | 113 | 30.6 | 54 | 149 |
| 5 | 13 | 126.5 | 144 | 24.9 | 85 | 149 |

**Table 5:** Descriptive Statistics

| Variable | N | Mean | Standard deviation | Min | Max |
|---|---|---|---|---|---|
| *Policy issue* | 69967 | 0.002 | 0.042 | 0 | 1 |
| *Ban* | 126 | 0.317 | 0.467 | 0 | 1 |
| *University rank* | 69967 | 259.493 | 143.731 | 1 | 501 |
| *Academic reputation* | 69967 | 39.617 | 26.431 | 2.6 | 100 |
| *Employer reputation* | 69967 | 39.267 | 28.679 | 1.5 | 100 |
| *Faculty student ratio* | 69967 | 49.250 | 32.510 | 1.5 | 100 |
| *Citation* | 69967 | 47.254 | 30.247 | 1 | 100 |
| *International faculty ratio* | 68658 | 45.188 | 39.634 | 1 | 100 |
| *International student ratio* | 69814 | 44.141 | 33.431 | 1.1 | 100 |
| *English speaking dummy* | 69967 | 0.336 | 0.472 | 0 | 1 |
| *Number of universities issuing policy* | 69472 | 0.525 | 0.993 | 0 | 3.738 |
| *Number of universities issuing banning policy* | 69472 | 0.303 | 0.600 | 0 | 2.302 |
| *ChatGPT sentiment* | 69967 | 50.690 | 27.156 | 2 | 100 |

Correlation Matrix

| | 1 | 2 | 3 | 4 | 5 | 6 | 7 | 8 | 9 | 10 | 11 | 12 |
|---|---|---|---|---|---|---|---|---|---|---|---|---|
| *Policy issue* | 1 | | | | | | | | | | | |
| *Ban* | NA | 1 | | | | | | | | | | |
| *University rank* | -0.024 | -0.204 | 1 | | | | | | | | | |
| *Academic reputation* | 0.024 | 0.178 | -0.839 | 1 | | | | | | | | |
| *Employer reputation* | 0.019 | 0.119 | -0.741 | 0.813 | 1 | | | | | | | |
| *Faculty-student ratio* | -0.003 | 0.232 | -0.302 | 0.127 | 0.136 | 1 | | | | | | |
| *Citation* | 0.020 | 0.217 | -0.467 | 0.326 | 0.217 | -0.281 | 1 | | | | | |
| *International faculty ratio* | 0.021 | 0.103 | -0.369 | 0.170 | 0.175 | -0.056 | 0.292 | 1 | | | | |
| *International student ratio* | 0.022 | 0.148 | -0.389 | 0.213 | 0.255 | 0.070 | 0.166 | 0.655 | 1 | | | |
| *English speaking dummy* | 0.035 | -0.186 | -0.159 | 0.117 | 0.115 | -0.158 | 0.256 | 0.392 | 0.413 | 1 | | |
| *Number of universities issuing policy* | 0.054 | -0.197 | -0.093 | 0.062 | 0.048 | -0.041 | 0.169 | 0.161 | 0.180 | 0.561 | 1 | |
| *Number of universities issuing banning policy* | 0.055 | -0.183 | -0.098 | 0.066 | 0.049 | -0.047 | 0.156 | 0.180 | 0.204 | 0.529 | 0.966 | 1 |
| *ChatGPT sentiment* | 0.034 | -0.055 | 0.046 | -0.042 | -0.031 | 0.004 | -0.039 | -0.035 | -0.040 | -0.061 | 0.416 | 0.404 |

Note: Number of universities issuing policy and number of universities issuing banning policy are not used in the same analysis.

**Table 6**: Model Analysis: Issuing a ChatGPT Policy

| Variable | Estimate | Standard Error | P-value |
|---|---:|---:|---:|
| *Intercept* | -9.022** | 1.059 | 0.000 |
| *University rank* | -0.001 | 0.002 | 0.491 |
| *Academic reputation* | 0.018** | 0.008 | 0.032 |
| *Employer reputation* | -0.007 | 0.007 | 0.269 |
| *Faculty/student ratio* | -0.005 | 0.004 | 0.192 |
| *Citation* | 0.003 | 0.005 | 0.591 |
| *International faculty ratio* | -0.008 | 0.005 | 0.128 |
| *International student ratio* | 0.000 | 0.005 | 0.994 |
| *English speaking* | 1.057* | 0.525 | 0.044 |
| *Number of universities issuing policy* | -0.129 | 0.150 | 0.390 |
| *ChatGPT sentiment* | 0.045*** | 0.006 | 0.000 |
| *Country fixed effects* | Included | | |
| *Log-likelihood* | -853.5 | | |

Note: *p < 0.1. **p < 0.05. ***p < 0.01. All two-tailed tests.

**Table 7**: Model Analysis: Ban or Not

| Variable | Estimate | Standard Error | P-value |
|---|---:|---:|---:|
| *Intercept* | -8.076 | 3.271 | 0.014 |
| *University rank* | 0.006 | 0.005 | 0.243 |
| *Academic reputation* | 0.016 | 0.024 | 0.523 |
| *Employer reputation* | -0.001 | 0.020 | 0.970 |
| *Faculty/student ratio* | 0.031** | 0.014 | 0.021 |
| *Citation* | 0.033* | 0.017 | 0.054 |
| *International faculty ratio* | -0.014 | 0.014 | 0.326 |
| *International student ratio* | -0.014 | 0.014 | 0.326 |
| *English speaking* | 2.985* | 1.607 | 0.063 |
| *Number of universities issuing banning policy* | -1.838** | 0.742 | 0.013 |
| *Generative AI sentiment* | 0.046 | 0.029 | 0.105 |
| *Country fixed effects* | *Included* | | |
| Log-likelihood | -47.5 | | |

Note: *p < 0.1. **p < 0.05. ***p < 0.01. All two-tailed tests.

**Figure 1A**: How Number of Universities Issued ChatGPT Policy Changes with the Number of Days Lapsed

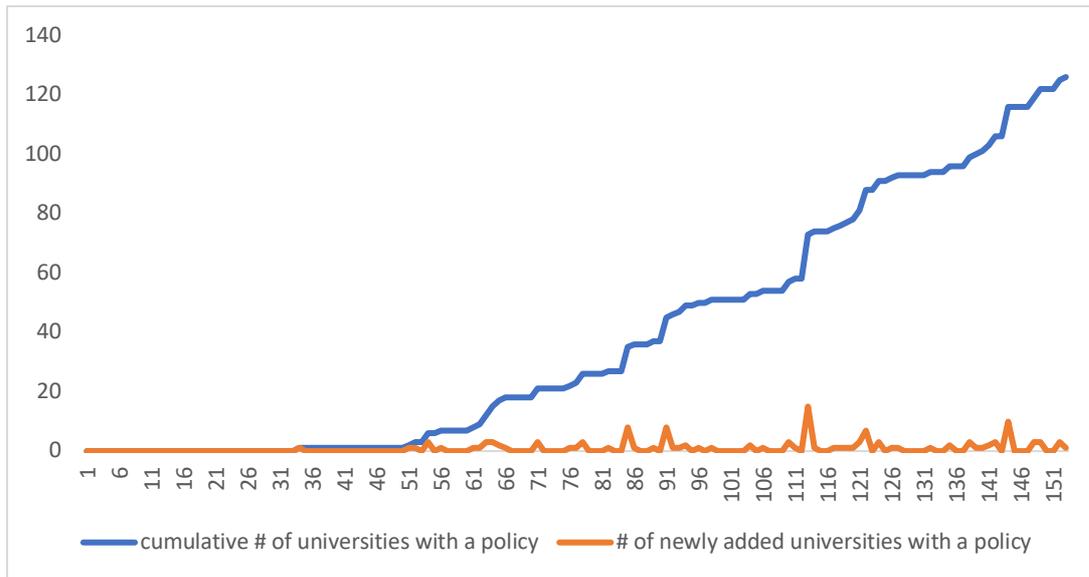

**Figure 2B:** How Number of Universities Issued Policy Banning or Embracing ChatGPT Changes with the Number of Days Lapsed

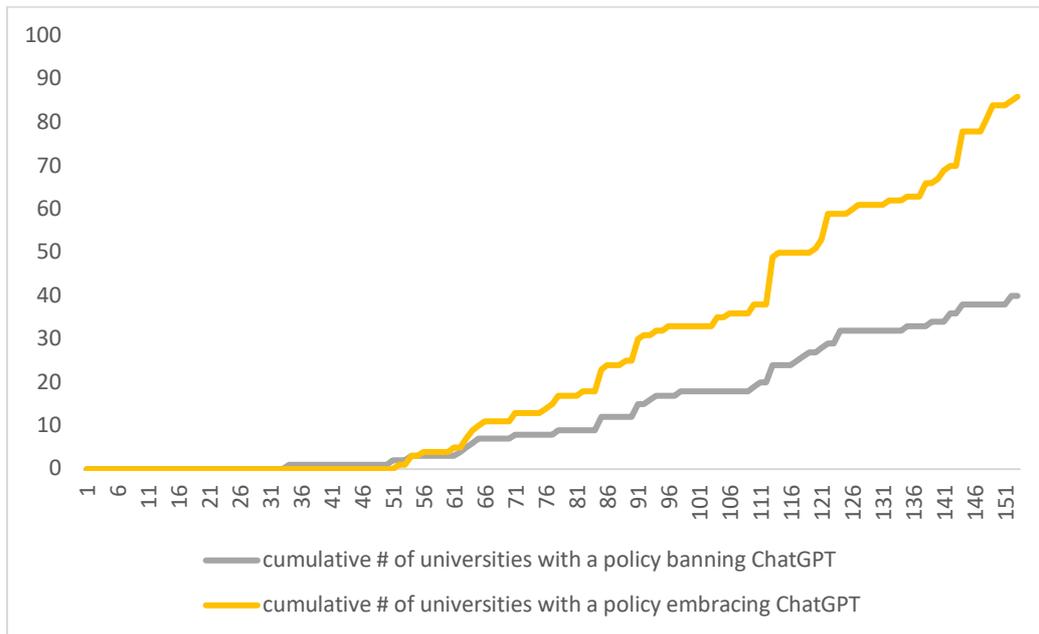

# Appendix

Table 1A: # of Universities in Each Country

| Country (Region) | # of Universities | Percent (%) | Country (Region) | # of Universities | Percent (%) |
|---|---|---|---|---|---|
| United States | 87 | 17.37 | Norway | 4 | 0.8 |
| United Kingdom | 49 | 9.78 | Portugal | 4 | 0.8 |
| Germany | 31 | 6.19 | Saudi Arabia | 4 | 0.8 |
| Australia | 26 | 5.19 | South Africa | 4 | 0.8 |
| China (Mainland) | 26 | 5.19 | Chile | 3 | 0.6 |
| Canada | 17 | 3.39 | Colombia | 3 | 0.6 |
| Russia | 17 | 3.39 | Czech Republic | 3 | 0.6 |
| Japan | 16 | 3.19 | Kazakhstan | 3 | 0.6 |
| South Korea | 16 | 3.19 | Pakistan | 3 | 0.6 |
| Italy | 14 | 2.79 | United Arab Emirates | 3 | 0.6 |
| Netherlands | 13 | 2.59 | Brunei | 2 | 0.4 |
| Spain | 12 | 2.4 | Iran, Islamic Republic of | 2 | 0.4 |
| France | 11 | 2.2 | Mexico | 2 | 0.4 |
| Taiwan | 10 | 2 | Poland | 2 | 0.4 |
| Switzerland | 9 | 1.8 | Singapore | 2 | 0.4 |
| Belgium | 8 | 1.6 | Thailand | 2 | 0.4 |
| India | 8 | 1.6 | Uruguay | 2 | 0.4 |
| Malaysia | 8 | 1.6 | Belarus | 1 | 0.2 |
| New Zealand | 8 | 1.6 | Cyprus | 1 | 0.2 |
| Sweden | 8 | 1.6 | Egypt | 1 | 0.2 |
| Finland | 7 | 1.4 | Estonia | 1 | 0.2 |
| Hong Kong SAR | 6 | 1.2 | Greece | 1 | 0.2 |
| Argentina | 5 | 1 | Lebanon | 1 | 0.2 |
| Austria | 5 | 1 | Lithuania | 1 | 0.2 |
| Brazil | 5 | 1 | Macau SAR | 1 | 0.2 |
| Denmark | 5 | 1 | Oman | 1 | 0.2 |
| Ireland | 5 | 1 | Peru | 1 | 0.2 |
| Israel | 5 | 1 | Philippines | 1 | 0.2 |
| Indonesia | 4 | 0.8 | Qatar | 1 | 0.2 |
| Norway | 4 | 0.8 | | | |